\documentclass[12pt]{article}
\UseRawInputEncoding
\begin{document}

\begin{center}
{\large \bf Brownian motion of neutrinos in the Sun 

[On the distribution of collisions of a neutrino with nucleons of the Sun]}
\vspace{0.5 cm}

\begin{small}
\renewcommand{\thefootnote}{*}
L.M.Slad\footnote{slad@theory.sinp.msu.ru} \\
{\it Skobeltsyn Institute of Nuclear Physics,
Lomonosov Moscow State University, Moscow 119991, Russia}
\end{small}
\end{center}

\vspace{0.3 cm}

\begin{footnotesize}
\noindent
{\bf Abstract.} When solving the solar neutrino problem on the basis of the hypothesis of the existence of a new interaction between electron neutrinos and nucleons, carried by a massless pseudoscalar boson, it becomes necessary to find the consequences of the short-term Brownian motion of neutrinos in the Sun. Earlier, we transmitted these consequences by the effective number of collisions of a neutrino with nucleons of the Sun, assuming the equality of fluxes near the Earth of left- and right-handed solar neutrinos. In this paper, we analyze the transfer of the mentioned consequences  by a test geometric distribution of collisions with one free parameter. This distribution provides good agreement between theoretical and experimental results for all five observed processes with solar neutrinos, while it gives the ratio of fluxes near the Earth of left- and right-handed neutrinos equal to 0.516:0.484. With such a ratio, the acceptability of the method of the effective number of collisions of a neutrino with nucleons is again confirmed. 
\end{footnotesize}

\vspace{0.5 cm}

\begin{small}

\begin{center}
{\large \bf 1. Introduction}
\end{center}

The problem of solar neutrinos has already arose as a result of the first experiment on their registration, namely, with the help of transitions $\nu_{e}+{}^{37}{\rm Cl} \rightarrow e^{-}+{}^{37}{\rm Ar}$ \cite{A}.  Davis and colleagues stated the upper limit for such a transition rate as 3 SNU (1 SNU is $10^{-36}$ captures per target atom per second), while the rate predicted by Bahcall \cite{B} was $30^{+30}_{-15}$ SNU.

Afterward, the emergence of the solar neutrino problem was assigned to the existence of neutrino oscillations in vacuum, a hypothesis of which was proposed by Gribov and Pontecorvo \cite{1}. The transformation of this hypothesis, attributed to neutrinos in the medium, was basically formulated by Wolfenstein \cite{2}, and subsequently received wide known as the MSM mechanism thanks to the Mikheev-Smirnov work \cite{3}.
 
Only recently, an alternative solution to the problem of solar neutrinos has been proposed which based on the hypothesis of the existence of a new, enough hidden fundamental interaction \cite{4}. 

In our work \cite{4}, we have already noted that, in spite of hundreds of papers devoted to the solar neutrino oscillations, there is yet no publication in the world literature in which the summarized theoretical results for the rates of all five processes observed with solar neutrinos would be presented in comparison with the corresponding experimental results. This assertion is confirmed, in particular, by the content of the regularly updated review of neutrino physics presented by Nakamura and Petcov in issues of the Particle Data Group \cite{C}. 

Let us note that the only analytical and logical tool available to the MSW mechanism is the Wolfelstein equation  written in the basis $(\nu_{e}, \nu_{\mu})$, as in \cite{C}, or in the basis $(\nu_{1}, \nu_{2})$, as in \cite{2}.
In our work \cite{5}, we pointed out the contradiction between the key assertions of Mikheev and Smirnov \cite{3} and the consequences of this equation. Moreover, we have proved that the MSW mechanism with parameters given by the SNO and Super-Kamiokande collaborations, taking into account the volume distribution of neutrino sources in the Sun, contradicts the results of three out of five observed processes with solar neutrinos.

The new interaction hypothesis \cite{4} is based on the logic and methods of classical quantum field theory. It leads to a change in the energy of the solar neutrino, allows a change in handedness and leaves its kind unchanged. The hypothesis assumes the existence of a massless pseudoscalar boson having a Yukawa coupling with electron neutrinos and nucleons, described by the following Lagrangian
\begin{equation}
{\cal L} = ig_{\nu_{e}ps}\bar{\nu}_{e}\gamma^{5}\nu_{e}\varphi_{ps}+
ig_{Nps}\bar{p}\gamma^{5}p\varphi_{ps}-ig_{Nps}\bar{n}\gamma^{5}n\varphi_{ps},
\label{1}
\end{equation}
and not interacting with electrons at the tree level.

The masslessness of the pseudoscalar boson ensures its stability, the impossibility of any decay.

We note three consequences of interaction (\ref{1}).

1. Since the pseudoscalar current connects the left and right components of the Dirac bispinor, then at each collision of an electron neutrino with a nucleon caused by the interaction (\ref{1}), the neutrino changes its handedness from left to right and vice versa. In the work \cite{4}, we assumed that the fluxes of left- and right-handed solar neutrinos near the Earth's surface are approximately equal.

2. The total cross section of the elastic scattering of an electron neutrino with energy $\omega$ by a nucleon at rest with mass $M$, given by the expression
\begin{equation}
\sigma = \frac{(g_{\nu_{e}ps}g_{Nps})^{2}}{16 \pi M^{2}}
\cdot \frac{1}{(1+2 \omega/M)},
\label{2}
\end{equation}
can be considered to be independent of the solar neutrino energy $\omega$, since 
$\omega \leq 18.8$ MeV \cite{6}.

3. At each collision of a neutrino with a nucleon of the Sun, its energy decreases. The energy of the scattered neutrino $\omega'$ can take evenly distributed values in the interval
\begin{equation}
\frac{\omega}{1+2\omega/M} \leq \omega' \leq \omega.
\label{3}
\end{equation} 

The task of accurately calculating the consequences of the short-term Brownian motion of each solar neutrino from the moment of its production to its exit from the Sun, which is caused by $\nu_{e} N$-collisions described by the Lagrangian (\ref{1}), is extremely difficult, if solvable at all. This task has only one free parameter, namely the constant $\beta \equiv g_{\nu_{e}ps}g_{Nps}/4\pi$. It should take into account the distribution of nucleons and neutrino sources over the volume of the Sun, given by the standard solar model \cite{6}. The main of the mentioned consequences is the distribution of $P_{\beta}(n)$ over the number $n$ of collisions of a neutrino with nucleons of the Sun.

In works \cite{4} and \cite{7}, this distribution is considered to be independent of the neutrino energy due to the relation (\ref {2}), as well as to be independent of the neutrino source. There were given the motives for replacing the unknown distribution $P_{\beta}(n)$ with the effective number of collisions, which serves as the only free parameter of the theory and can be both integer $ n_ {0} $, and non-integer $n_{a}$ ($n_{0} <n_{a} <n_{0} + 1$). The number $n_{0} $ describes the final energy distribution of a neutrino with a fixed initial energy. In the work \cite{4}, it is shown that a good description of all five processes with solar neutrinos is provided when the value of the effective number of collisions is 11.  

In this work, we use, as a trial distribution of collisions of a neutrino with nucleons that occurred from the moment of neutrino production to its exit from the Sun, the geometric distribution well-known in mathematics
\begin{equation}
P_{\beta}(n) = p(1-p)^{n}, \qquad n=0,1,2, \ldots,
\label{4}
\end{equation}
where $p$ is a free parameter. 

Choosing a trial distribution $P(n)$ among discrete distributions well known in mathematics, we were guided by plausible requirements that the probability of a neutrino leaving the Sun without collisions with nucleons in it be nonzero and that the sequence $P(n)$ be monotonically decreasing with increasing $n$. The geometric distribution is the only one among all that satisfies these requirements. Since the solar environment is described by spherically symmetric volume distributions of neutrino sources, matter, atomic nuclei, etc., then without solving the problem of mathematically accurate finding of the consequences of the Brownian motion of neutrinos in the Sun, there is no way to find any connection between the parameter $p$ in (\ref{4}) with such solar distributions.

\begin{center}
{\large \bf 2. Geometric distribution of collisions of a neutrino with nucleons of the Sun}
\end{center}

The results of theoretical calculations of the rates of all observed processes with solar neutrinos, based on the Lagrangian (\ref{1}) and on the distribution (\ref{4}), are in the best agreement with the experimental results if $p = 0.062$.

The probability $W_{\rm left}$ that a solar neutrino near the Earth's surface is left-handed is the probability that, while moving in the Sun, a neutrino has experienced an even number of collisions with nucleons. It is equal
\begin{equation}
W_{\rm left} = \sum_{k=0}^{\infty} P_{\beta}(2k) = \sum_{k=0}^{\infty}p(1-p)^{2k} =
(2-p)^{-1} = 0.516.
\label{5}
\end{equation}
The given value of the probability $W_{\rm left}$ introduces a small correction to our initial assumption, expressed in the work \cite{4}, about the approximate equality of the fluxes of left- and right-handed solar neutrinos near the Earth's surface. We have supplemented the calculations of the rates of processes (subprocesses) caused by left-handed neutrinos, performed in the work \cite{4}, with calculations with their fluxes $0.516 \Phi$ instead of $0.5 \Phi$, and we present new theoretical results below.

When calculating the rates of the observed processes, taking into account the distribution of collisions of a neutrino with nucleons of the Sun (\ref{4}), we, just as in the work \cite{4}, assume that as a result of each collision, the neutrino energy with equal probability acquires one of two boundary values of the interval (\ref{3}) and thus after $n$ collisions the initial energy level $\omega$ turns into a set of $n + 1$ binomially distributed values.

Let $v_{A}(n)$ denote the rate of the process (subprocess) $A$, which is caused by neutrinos with an energy spectrum appearing after $n$ collisions of each neutrino with solar nucleons due to the interaction (\ref{1}). Then the rate $V_{A}$ of the process (subprocess) $A$ in a terrestrial installation, generated by left-handed solar neutrinos, taking into account the geometric distribution of collisions in the Sun (\ref{4}), is found by the following formula
\begin{equation}
V_{A} = \sum_{k=0}^{\infty} P_{\beta}(2k) v_{A}(2k) = \sum_{k=0}^{\infty}p(1-p)^
{2k} v_{A}(2k). 
\label{6}
\end{equation}

The formula (\ref{6}) refers to all considered processes (subprocesses), except for the subprocess of deuteron disintegration, caused by the exchange with a pseudoscalar boson $\varphi_{ps}$ between any solar neutrinos and deuteron nucleons. The rate of this subprocess is given by the expression
\begin{equation}
V_{A_{\varphi}} = \sum_{n=0}^{\infty} P_{\beta}(n) v_{A_{\varphi}}(n) = \sum_{n=0}^{\infty}p(1-p)^{n} v_{A_{\varphi}}(n). 
\label{7}
\end{equation}

The following formula for the transitions $\nu_{e}+{}^{37}{\rm Cl} \rightarrow e^{-}+{}^{37}{\rm Ar}$ can serve as an example of calculating the rate of the process $A$,  caused by neutrinos from ${}^{8}{\rm B}$ decays,
\begin{equation}
V({}^{37}{\rm Cl} \ | \ {\rm B}) = \sum_{k=0}^{\infty} p(1-p)^{2k} 
\Phi({}^{8}{\rm B})\sum_{i=1}^{160}\Delta^{B}p(\omega_{i}^{B})
\sum_{n=1}^{2k+1}\frac{(2k)!}{2^{2k}(n-1)!(2k+1-n)!}
\sigma^{\rm Cl}(\omega_{n,i}^{B}), 
\label{8}
\end{equation}
where $\Phi({}^{8}{\rm B})$ is the total neutrino flux from ${}^{8}{\rm B}$, ${}^{8}{\rm B} = 5.79 \times 10^{6}(1 \pm 0.23)$ \cite{8}; neutrino energy values ​​are given as the set $\omega_{i}^{B}=i\Delta^{B}$, где $i = 1, \ldots, 160$, $\Delta^{B}=0.1$ MeV, and their distribution given in \cite{9} is expressed in terms of the probability $p(\omega_{i}^{B})$ that the neutrino has energy in the interval$(\omega_{i}^{B}-\Delta^{B}/2, \ \omega_{i}^{B}+\Delta^{B}/2)$; $\sigma^{\rm Cl}(\omega)$ is the cross section of the absorption process by chlorine of neutrinos with energy $\omega$ \cite {6}; $\omega_{n,i}^{B}$ is the neutrino energy acquired after $n-1$ collisions with nucleons, at its initial value $\omega_{i}$.

The information needed to calculate the rates of other observed processes is available in the work \cite{4}.

Tables 1-5 below show the results of calculating the velocities of all observed processes with solar neutrinos based on the method of the effective number of neutrino-nucleon collisions $n_{0}$ for two values of the probability $W_{\rm left} $ that at the Earth's surface the solar neutrino is left-handed, and based on geometric distribution (\ref {4}). Tables 3-4 show the dependence of the rates of the processes on the lower value $E_{c}$ of the reconstructed energy of the final electron.

\begin{center}
\begin{tabular}{lccccccc}
\multicolumn{8}{c}{{\bf Table 1.} The rate of transitions ${}^{37}{\rm Cl} \rightarrow {}^{37}{\rm Ar}$ in SNU.} \\
\hline
\multicolumn{1}{l}{} 
&\multicolumn{1}{c}{${}^{8}{\rm B}$} 
&\multicolumn{1}{c}{${}^{7}{\rm Be}$}
&\multicolumn{1}{c}{${}^{15}{\rm O}$}
&\multicolumn{1}{c}{$pep$}
&\multicolumn{1}{c}{${}^{13}{\rm N}$}
&\multicolumn{1}{c}{$hep$}
&\multicolumn{1}{c}{Total} \\ 
\hline
Experiment \cite{10} &  &  &  &  &  &  & $2.56 \pm 0.16 \pm 0.16$ \\
$W_{\rm left}=0.5$, $n_{0} = 11$ & 1.97 & 0.43 & 0.17 & 0.11 & 0.04 & 0.01 & 2.72 \\
$W_{\rm left}=0.516$, $n_{0} = 12$ & 1.95 & 0.43 & 0.17 & 0.11 & 0.05 & 0.01 & 2.72 \\
Eq. (\ref{4}), $p=0.062$ & 2.02 & 0.42 & 0.17 & 0.11 & 0.04 & 0.01 & 2.77 \\
\hline
\end{tabular}
\end{center}

\begin{center}
\begin{tabular}{lcccccccc}
\multicolumn{9}{c}{{\bf Table 2.} The rate of transitions ${}^{71}{\rm Ga} \rightarrow {}^{71}{\rm Ge}$ in SNU.} \\ 
\hline
\multicolumn{1}{l}{}
&\multicolumn{1}{c}{$p$-$p$}
&\multicolumn{1}{c}{${}^{7}{\rm Be}$} 
&\multicolumn{1}{c}{${}^{8}{\rm B}$} 
&\multicolumn{1}{c}{${}^{15}{\rm O}$}
&\multicolumn{1}{c}{${}^{13}{\rm N}$}
&\multicolumn{1}{c}{$pep$}
&\multicolumn{1}{c}{$hep$}
&\multicolumn{1}{c}{Total} \\ 
\hline
Experiment \cite{11} &  &  &  &  &  &  &  & $62.9^{+6.0}_{-5.9}$ \\
Experiment \cite{12} &  &  &  &  &  &  &  & $65.4^{+3.1}_{-3.0}{}^{+2.6}_{-2.8}$ \\
$W_{\rm left}=0.5$, $n_{0} = 11$ & 34.6 & 17.2 & 4.9 & 2.8 & 1.7 & 1.4 & 0.02 & 62.6 \\
$W_{\rm left}=0.516$, $n_{0} = 12$ & 35.7 & 17.7 & 4.9 & 2.9 & 1.7 & 1.4 & 0.02 & 64.4 \\
Eq. (\ref{4}), $p=0.062$ & 35.6 & 17.6 & 5.0 & 2.8 & 1.7 & 1.4 & 0.02 & 64.2 \\
\hline
\end{tabular}
\end{center}

\begin{center}
{{\bf Table 3.} Effective fluxes of neutrinos $\Phi_{eff}^{\nu e}({}^{8}{\rm B})$ found from the process} \\
\begin{tabular}{lccccc}
\multicolumn{6}{l}{$\nu_{e} e^{-}\rightarrow \nu_{e} e^{-}$ ($E_{c}$ is given in MeV, and the fluxes are in units of $10^{6}$ ${\rm cm}^{-2}{\rm s}^{-1}$).} \\  
\hline
\multicolumn{1}{l}{References}
&\multicolumn{1}{c}{$E_{c}$} 
&\multicolumn{1}{c}{Experimental}
&\multicolumn{1}{c}{$W_{\rm left}=0.5,$} 
&\multicolumn{1}{c}{$W_{\rm left}=0.516,$}
&\multicolumn{1}{c}{Eq. (\ref{6}),} \\
&& results & {$n_{0} = 11$} & {$n_{0} = 12$} & {$p=0.062$} \\
\hline
SK III \cite{13} & 5.0 & $2.32\pm 0.04 \pm 0.05$ & 2.27 & 2.29 & 2.26 \\
SNO I \cite{14} & 5.5 &$2.39^{+0.24}_{-0.23}{}^{+0.12}_{-0.12}$ & 2.19 & 2.20 & 2.17\\
SNO II \cite{15} & 6.0 &$2.35\pm 0.22\pm 0.15$ & 2.10 & 2.10 & 2.09\\
SNO III \cite{16} & 6.5 &$1.77^{+0.24}_{-0.21}{}^{+0.09}_{-0.10}$ & 2.01 & 2.01 & 2.01\\
\hline
\end{tabular}
\end{center}

\begin{center}
{{\bf Table 4.} Effective fluxes of neutrinos $\Phi_{eff}^{cc}({}^{8}{\rm B})$ found from the process} \\
\begin{tabular}{lccccc}
\multicolumn{6}{l}{$\nu_{e}D \rightarrow  e^{-}pp$ ($E_{c}$ is given in MeV,and the fluxes are in units of $10^{6}$ ${\rm cm}^{-2}{\rm s}^{-1}$).} \\ 
\hline
\multicolumn{1}{l}{References}
&\multicolumn{1}{c}{$E_{c}$} 
&\multicolumn{1}{c}{Experimental}
&\multicolumn{1}{c}{$W_{\rm left}=0.5,$} 
&\multicolumn{1}{c}{$W_{\rm left}=0.516,$}
&\multicolumn{1}{c}{Eq. (\ref{6}),} \\
&& results & {$n_{0} = 11$} & {$n_{0} = 12$} & {$p=0.062$} \\
\hline
SNO I \cite{14} & 5.5 & $1.76^{+0.06}_{-0.05}{}^{+0.09}_{-0.09}$ & 1.86 & 1.85 & 1.88 \\
SNO II \cite{15} & 6.0 & $1.68^{+0.06}_{-0.06}{}^{+0.08}_{-0.09}$ & 1.77 & 1.74 & 1.80 \\
SNO III \cite{16} & 6.5 & $1.67^{+0.05}_{-0.04}{}^{+0.07}_{-0.08}$ & 1.66 & 1.62 & 1.72 \\
\hline
\end{tabular}
\end{center}

\newpage

\begin{center}
{{\bf Table 5.} Effective fluxes of neutrinos $\Phi_{eff}^{nc}({}^{8}{\rm B})$ found from the process} \\
\begin{tabular}{lccccc}
\multicolumn{6}{c}{$\nu_{e}D \rightarrow  \nu_{e}np$ (the fluxes are in units of $10^{6}$ ${\rm cm}^{-2}{\rm s}^{-1}$).} \\
\hline
\multicolumn{1}{l}{}
&\multicolumn{1}{c}{Exchange}
&\multicolumn{1}{c}{Exchange}
&\multicolumn{1}{c}{Exchange} 
&\multicolumn{1}{c}{Exchange}
&\multicolumn{1}{c}{Sum} \\
& by $Z$ & by $Z$ & by $Z$ & by $\varphi$ & \\
& $W_{\rm left}=0.5$ & $W_{\rm left}=0.516$ & Eq. (\ref{6}) & all neutrinos & \\
\hline
SNO I \cite{14} & & & & & $5.09^{+0.44}_{-0.43}{}^{+0.46}_{-0.43}$ \\
SNO II \cite{15} & & & & & $4.94^{+0.21}_{-0.21}{}^{+0.38}_{-0.34}$  \\
SNO III \cite{16} & & & & & $5.54^{+0.33}_{-0.31}{}^{+0.36}_{-0.34}$ \\
\hline
$n_{0} = 11$ & 2.10 & & & 2.87 & 4.98 \\
$n_{0} = 12$ & & 2.11 & & 2.85 & 4.96 \\
Eq. (\ref{6})-(\ref{7}) & & & 2.13 & 2.80 & 4.92 \\
\hline
\end{tabular}
\end{center}

Comparing the corresponding numbers in each of the tables, we come to the following conclusion. The results of collisions of neutrinos with nucleons in the Sun caused by the interaction (\ref{1}), which were obtained on the basis of the method of the effective number of collisions both with equal fluxes of left- and right-handed neutrinos near the Earth's surface, and with a slight superiority of the flux of left-handed neutrinos, and which were obtained based on the geometric distribution of collisions, are close to each other and agree well with the experimental results. The most wispy agreement with experiments is for the effective number of collisions $ n_ {0} = 12 $ with a fraction of left-handed neutrinos of 0.516.

Note also that the geometric distribution (\ref{4}) corresponds to the following average number of collisions of a neutrino with nucleons in the Sun
\begin{equation}
\bar{n} = \sum_{n=0}^{\infty} nP_{\beta}(n) = p\sum_{n=0}^{\infty}n(1-p)^{n} =
p^{-1} = 16.1.
\label{9}
\end{equation}

\begin{center}
{\large \bf 3. Concluding remarks}
\end{center}

Having performed suitable calculations and comparisons, we recognize the fact that the geometric distribution can serve as an acceptable approximation of the real distribution of collisions of neutrinos with nucleons of the Sun, since it provides a good level of agreement between theoretical and experimental results. One of the logical consequences of this distribution is a very important ratio of the fluxes of left- and right-handed solar neutrinos near the Earth's surface. We can now say that an alternative solution to the solar neutrino problem has taken on a complete form.

\end{small}
\end{document}